# THE LIGO SUSPENDED OPTIC DIGITAL CONTROL SYSTEM

J. Heefner, R. Bork, LIGO Laboratory, California Institute of Technology, Pasadena, CA 91125, USA

Abstract

The original LIGO[1] suspension control system [1] used analog circuitry to implement the closed loop damping required for local control of each the suspended optics. Recent developments in analog to digital converters (ADC), digital to analog converters (DAC), increased processor speed and performance, and the use of reflective memory have made a digital alternative possible. Such a control system would provide additional performance and flexibility that will be required for operation of the interferometers. This paper will describe the real-time digital servo control systems that have been designed, developed and implemented for the LIGO suspended optics. In addition, the paper will describe how the suspension controls have been integrated into the overall LIGO control and data acquisition systems [2].

## 1 INTRODUCTION

The LIGO interferometers located in Hanford, Washington and Livingston, Louisiana are Michelson laser interferometers enhanced by multiple coupled optical resonators. These coupled optical resonators are 4 kilometer long Fabry-Perot cavities placed in each arm of the interferometer. The mirrors that form the cavities are suspended from a single loop of wire mounted inside suspension cages that are, in turn, mounted on seismically isolated optical platforms within the LIGO vacuum system. Control of the optic is achieved using voice coil actuators that act on magnets attached to the surface of each optic. Shadow sensors are used to measure movement and orientation of the optic with respect to the suspension cage.

There are two types of optic suspensions in LIGO. The Large Optic Suspension and the Small Optic Suspension. Generically these suspension types are identical and only differ in optic size, dynamic range of motion and absolute noise floor of the controls. In the case of the Large Optic Suspension controls the key requirements are:

- *Dynamic range of longitudinal motion: 20um$_{p-p}$*
- *Dynamic range of angular motion: 500urad$_{p-p}$*
- *Noise: 5 x 10$^{-20}$*(f/40)$^{-2}$ for f>40Hz*

Key requirements of the Small Optic controls are:

- *Dynamic range of longitudinal motion: 27um$_{p-p}$*
- *Dynamic range of angular motion: 1500urad$_{p-p}$*
- *Noise: 3 x 10$^{-18}$*(f/40)$^{-2}$ for f>40Hz*

The Large and Small Optic controls have two functional requirements:

- Provide for "local" damping of the suspended optic using the shadow sensors and voice coil actuators. (Nominally 6 Large and 7 Small).
- Provide a means for the LIGO Length (LSC) and Alignment (ASC) controls to control the longitudinal position and orientation of the optic.

## 2 HARDWARE AND SYSTEM DESIGN

In the original LIGO suspension control system, the local damping loop for each optic was implemented using a completely analog solution. In this the solution, the shadow sensor signals were amplified and combined to form signals corresponding to longitudinal position, pitch and yaw readings. A 10-pole chebychev filter with a zero at DC then filtered these signals. The filtered signals were then combined with inputs from the LSC and ASC to form control signals in the position, pitch and yaw degrees of freedom. These signals were passed to an output matrix and sent to each of the actuators. Early on in the commissioning of the LIGO interferometers it was recognized that a digital solution would provide the additional performance and flexibility required for operation of the interferometers at their design sensitivity. Some of the limitations of the analog option were:

- The velocity damping provided by the chebychev filter was not necessarily optimal and in some cases a DC coupled servo was desired. In other cases it was desired to have different filter characteristics for different optics or different modes of interferometer operation. These characteristics included gain bubbles at the micro-seismic peak (~0.16 Hz).
- The input and output matrices were simple gain stages. A more optimal solution required frequency shaping in the matrices.
- The interface to the LSC and ASC systems was via cables and connectors where the ASC and LSC signals originate in racks many meters away. Interference and noise was an issue for these sensitive signals.

---
[1] Laser Interferometer Gravitational-Wave Observatory

- The ASC and LSC systems [3] were implemented using digital servos and a more integrated approach including the suspension systems was desired.

After carefully considering the alternatives and the operational issues involved with replacing existing systems it was decided to implement a digital alternative for the suspension controls. This solution uses VME based CPUs, ADCs, DACs and reflective memory modules. The models and types are listed in Table 1.

Table 1: VME modules used in systems

| Type | Model | Characterisitcs |
| --- | --- | --- |
| CPU | VMIC[2] Pentium III | 850MHz to 1GHZ, VxWorks® OS |
| ADC | ICS[3]-110B | 16 bit, 32 Channels |
| DAC | Pentek[4] 6102 | 16 bit, 8 channels |
| Reflective Memory | VMIC 5579 | PMCbus, single mode and multi-mode fiber |

All custom analog electronics used for signal conditioning and actuator drive are mounted in 19-inch rack mount chassis or Eurocard format crates located in the suspension racks. These custom electronics are typically very low noise (1-2nV/√Hz), high dynamic range circuits. The timing for the ADCs and DACs is derived from the same GPS (Global Positioning System) based timing system used by the LIGO data acquisition system [2].

Since the suspension controls act as the actuation point for the ASC and LSC systems, the suspension controls had to be integrated into their reflective memory networks. There are three reflective memory loops for the ASC and LSC controls, one for each arm of the interferometer and one for the corner station. In addition there is a reflective memory network for the LIGO data acquisition system. Each node of the network is connected in a loop using single mode or multi-mode fiber. Data written to a memory location at one node is passed to each node in the loop. In this way data can be shared among nodes.

## 3 LOCAL DAMPING LOOP

All of the software is modular and built around a block of code that contains gain, 3 filters (up to 4 second order sections each), an enable, offset adjust and an excitation input. Each of the 3 filters can be enabled or bypassed on the fly by the operator. Figure 1 shows the structure of a basic code building block (BCBB).

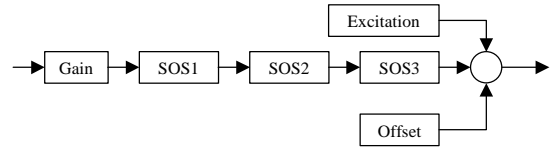

Figure 1: Basic Code Building Block (BCBB)

All filters are implemented using IIR filters formed by second order sections (SOS1, SOS2, SOS3). The coefficients for each filter are stored as a text file that is read by the code on start up. This allows the operator to select different servo filters on the fly by enabling or bypassing various filters, or editing the filter text file and restarting the system can change the entire configuration.

The basic code blocks are then combined to form the algorithms for each of the suspension local damping loops. The figure below is a block diagram of the code necessary to form the local damping loop for a single optic.

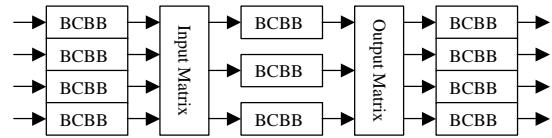

Figure 2: Block Diagram of Code For A Single Local Damping Loop

In the local damping loop for each optic, the four sensor inputs are combined to form the three degrees of freedom: position, pitch and yaw. These are shown top to bottom, respectively between the input and output matrices in Figure 2. The function of the output matrix is to form the four actuator signals from the three degrees of freedom. It should be noted that there is also a transverse degree of freedom for each optic that is not shown in the figure. This degree of freedom is handled separately by a single input, single output control loop formed by basic code building blocks. Not shown in the diagram is the addition of the length and alignment signals supplied by the LSC and ASC systems. These signals are added to their respective degrees of freedom just prior to the output matrix.

---

[2] VMIC, Huntsville, AL, USA
[3] Interactive Circuits and Systems, Ltd., Gloucester, Ontario, Canada
[4] Pentek, Inc., Upper Saddle River, NJ, USA

## 4 SYSTEM ARCHITECTURE

One of the functions of the suspension controls is to provide the LSC and ASC actuation on each of the mirrors. This function requires that the suspension controls be distributed throughout the LIGO site and also that they be tightly coupled to the LSC and ASC systems. This is accomplished by using multiple CPUs and VME crates coupled via fiber optic reflective memory loops. These reflective memory loops allow sharing of data from one processor to the next. The higher bandwidth requirements of the LSC system drive the clock rates for the actuator outputs to 16384 samples per second, far higher than would be required for the local damping loop alone. For these reasons, local controls were distributed among five VME crates as shown in Table 2.

Table 2: VME Crate Locations and Functions

| Location | Clock Rate | Functions |
|---|---|---|
| Vertex Crate 1 | 2048 | Read sensor inputs for all vertex optics<br>Perform input matrix and servo filters calcs for all vertex optics<br>Perform output matrix calcs and output actuator signals for all SOS<br>Perform mode cleaner length controls<br>DAQ/GDS interface for all suspensions |
| Vertex Crate 2 | 16384 | Output matrix calcs for up to three vertex LOS<br>LSC and ASC interface for up to three vertex LOS |
| Vertex Crate 3 | 16384 | Same as output crate 1 |
| X End | 16384 | In and Out matrix calcs for X end ASC X end functions |
| Y End | 16384 | Same as X end, but Y end |

The vertex input crate performs the equivalent of 750 second order section IIR filter calculations 2048 times per second and each of the output crates perform the equivalent of up to 80 sections at 16384 times per second.

## 5 INTERFACE TO DAQ, GDS AND EPICS

The LIGO DAQ and GDS systems are interfaced via reflective memory. One of the functions of the vertex input crate is to read and write data from and to this reflective memory. Using the DAQ and GDS an operator can look at data from the system, measure transfer functions, plot power spectra and input diagnostic test signals, in real time. A more complete description of the functions and operation of the DAQ and GDS systems are described in reference [2].

Routine operator interface is via EPICS[5]. Through EPICS, the operator can monitor the system, change matrices, gains and offsets, engage filters and invert servo polarity. A Motorola MVME 162 CPU is located in each of the VME crates listed in Table 2. This CPU processes the database records, state code, etc. for the controls located in that crate. Data is communicated between the front end CPU (Pentium III) and the EPICS CPU via shared memory across the VME backplane. This is done in an effort to minimize the burden on the front end CPU and allow it to handle only the time critical front end servo controls.

## 6 EXPERIENCE TO DATE AND FUTURE PLANS

At this time the digital suspension controls for the 4 kilometer interferometer located in Hanford have been installed and tested and the interferometer is being commissioned. All components for the other two interferometers have been prepared and are ready for installation. To date, no major changes to the basic design are anticipated and the design appears to meet all of the system requirements necessary for the interferometers to reach their full sensitivity.

## 7 ACKNOWLEDGEMENTS

We thank the entire LIGO team for assistance and support. This work was supported by National Science Foundation Grant PHY-920038.

---

[5] EPICS- Experimental Physics and Industrial Control System